\documentstyle[12pt]{article} \textheight22.0cm
\catcode `\@=11 \@addtoreset{equation}{section}

\def\section{\@startsection {section}{1}{\z@}{-3.5ex plus -1ex minus
-.2ex}{2.3ex plus .2ex}{\normalsize\bf}}
\def\subsection{\@startsection{subsection}{2}{\z@}{-3.25ex plus -1ex
minus -.2ex}{1.5ex plus .2ex}{\normalsize\bf}} 

\def\thebibliography#1{\section*{References\markboth
  {REFERENCES}{REFERENCES}}\list
  {[\arabic{enumi}]}{\settowidth\labelwidth{[#1]}\leftmargin\labelwidth
  \advance\leftmargin\labelsep \usecounter{enumi}}
  \def\newblock{\hskip .11em plus .33em minus -.07em} \sloppy
  \sfcode`\.=1000\relax}  \catcode
  `\@=12 \input{amssym.def} \input{amssym.tex} \newcommand{\x}{\times}
  \newcommand{\til}{\tilde} \newcommand{\Ad}{{\rm Ad}}
  \newtheorem{Definition}{Definition} \newtheorem{Theorem}{Theorem}
  \renewcommand{\ll}{\label} \newcommand{\be}{\begin{equation}}
  \newcommand{\ee}{\end{equation}} \newcommand{\bib}{\bibitem}
  \newcommand{\ci}{\cite} \newcommand{\ca}{$C^*$-algebra}
  \newcommand{\rep}{representation}  \newcommand{\Hs}{Hilbert space}
  \newcommand{\mom}{momentum map} \newcommand{\MW}{Marsden--Weinstein}
  \newcommand{\YM}{Yang--Mills} 
  \newcommand{\raw}{\rightarrow} \newcommand{\rat}{\mapsto}
  \newcommand{\n}{\|} \newcommand{\ot}{\otimes}
   \newcommand{\la}{\langle}
  \newcommand{\ra}{\rangle} \newcommand{\cin}{C^{\infty}}
  \newcommand{\half}{\mbox{\footnotesize $\frac{1}{2}$}}
  \newcommand{\Hh}{{\cal H}_{\hbar}} \newcommand{\PHh}{{\Bbb P}{\cal
  H}_{\hbar}} 
  \newcommand{\inv}{^{-1}} \newcommand{\Exp}{{\rm Exp}}

  \newcommand{\dl}{\delta} \newcommand{\Dl}{\Delta}
   
   \newcommand{\et}{\eta}
  \newcommand{\th}{\theta}

  \newcommand{\rh}{\rho} \newcommand{\sg}{\sigma}
   
   \newcommand{\Ph}{\Phi}
   \newcommand{\ch}{\chi}
   \newcommand{\Ps}{\Psi}
   \newcommand{\Om}{\Omega}
 \newcommand{\g}{{\frak g}}
 \newcommand{\k}{{\frak k}}

\newcommand{\CA}{{\cal A}} 
\newcommand{\CD}{{\cal D}} \newcommand{\CE}{{\cal E}}
\newcommand{\CG}{{\cal G}} \renewcommand{\H}{{\cal H}}
 \newcommand{\CI}{{\cal I}}
\newcommand{\CK}{{\cal K}} 
\newcommand{\CN}{{\cal N}} 
 \newcommand{\CP}{{\cal P}}
\newcommand{\CQ}{{\cal Q}} 
 \newcommand{\CV}{{\cal V}}
\newcommand{\CW}{{\cal W}} 

\newcommand{\C}{{\Bbb C}} 
 \newcommand{\I}{{\Bbb I}}
 \newcommand{\R}{{\Bbb R}}
\newcommand{\T}{{\Bbb T}} 
\begin{document}
\vspace*{2.5cm}
\noindent
{ \bf HALL'S COHERENT STATES, THE CAMERON--MARTIN THEOREM, AND THE
QUANTIZATION OF YANG--MILLS THEORY ON A CIRCLE\footnote{Submitted to the
Proceedings of the  XVIIth
Workshop on Geometric Methods in Physics, Bia\l owie\.{z}a, 1998,,
eds.\  M. Schlichenmaier, S.T. Ali, and A. Strasburger} }\vspace{1.3cm}\\
\noindent
\hspace*{1in}
\begin{minipage}{13cm}
N.P. Landsman$^{1}$ and K.K. Wren$^{2}$ \vspace{0.3cm}\\ $^{1}$
 Korteweg--de Vries Institute for Mathematics\\ University of
 Amsterdam\\ Plantage Muidergracht 24, 1018 TV AMSTERDAM\\ THE
 NETHERLANDS \\ \makebox[3mm]{ }E-mail: npl@wins.uva.nl \vspace{0.1cm}
 \\ $^{2}$ Cazenove \& Co., 12 Tokenhouse Yard\\ LONDON EC2R 7AN,
 U.K.\\ \makebox[3mm]{ }E-mail: kkwren@cazenove.com
\end{minipage}
\vspace*{0.5cm}

\begin{abstract}
\noindent
We discuss the classical and quantum reduction to the space of
physical degrees of freedom of Yang--Mills theory on a circle (so that
space-time is a cylinder).  Although the classical reduced phase space
is finite-dimensional, the quantum reduction procedure is
mathematically fascinating, involving firstly the Wiener measure on a
loop group, secondly a generalization of the Cameron--Martin theorem
to loop groups, and thirdly Hall's coherent states for compact Lie
groups. Our approach is based on a quantum analogue of the classical
Marsden--Weinstein symplectic reduction process.
\end{abstract}
\section{\hspace{-4mm}.\hspace{2mm}INTRODUCTION}
Yang--Mills theories on a two-dimensional space-time serve as a
laboratory for studying issues that are relevant to general gauge
theories. Even without fermions, one may look at the role of gauge
invariance, constraints, reduction, observables, the geometry of the
space of physical degrees of freedom, possible singularities occurring
in the latter, and what not.  Here it is of particular interest to
understand how the structure of the classical theory is reflected in
the quantum theory.  If one includes fermions, one may in addition
look at anomalies, spectral flow, supersymmetry, etc. Moreover, there
are unexpected connections to string theory and $M$-theory
\ci{Moo,DVV}.

The advantage of the two-dimensionality of the model is that, with the
present state of the art, one may proceed in a mathematically rigorous
fashion.  The fact that the model is physically somewhat trivial then
turns out to be compensated by an astonishingly rich mathematical
structure. On the geometry side, this seems to have been first
realized by Witten \ci{Wit1,Wit2}, who considered the Euclidean
version in which the the theory is defined on a Riemann surface (also
cf.\ \ci{Sen1,Sen2}). As we shall see, even the much simpler Minkowski
formulation on a circle, in which space-time is taken to be a
cylinder, involves matters of a certain interest to analysis and
measure theory \ci{LW,Wren2,Wren3,MT}.

In what follows, we shall always speak about this particular version
of Yang--Mills theory. Without loss we will work in the temporal gauge
$A_0=0$, in which the residual gauge transformations are
time-independent. Thus the gauge group consists of loops in the
structure group $K$.  Since in this paper we will not discuss the
singular structure of the physical theory (cf.\ \ci{Wren1,Wren3}), we
further specialize to the group of based gauge transformations, which
consists of loops in $K$ that start and end at the unit element $e$.

Our central concern in this paper is the reduction from the degrees of
freedom that are originally given (viz.\ the space of all gauge fields
and their conjugate momenta) to the space of physical degrees of
freedom.  As first remarked by Rajeev \ci{Raj1}, the reduced phase
space of the classical theory is finite-dimensional, being canonically
isomorphic to the cotangent bundle $T^*K$.  As in four-dimensional
Yang--Mills theory \ci{Arms}, one may see this reduction as a special
instance of a Marsden--Weinstein quotient in symplectic geometry
\ci{AM,MT}.  While this neat geometric fact does not dramatically
change the perspective on the classical theory, it lies at the basis
of our quantization procedure. For we are going to quantize the theory
by first quantizing the unconstrained phase space, and subsequently
implementing a recent proposal \ci{JGP,bial} to quantize the
Marsden--Weinstein reduction process by a technique based on the
theory of induced \rep s of \ca s \ci{MT} (for different rigorous
approaches see \ci{Dim,ALMMT,HD}).

It turns out that this technique can be carried through in the case at
hand, involving an integration over the gauge group. Now, two
interesting things happen.  Firstly, the physical theory turns out to
be gauge invariant as a consequence of the Cameron-Martin theorem for
loop groups \ci{Fre,MM}.  We will provide an interesting perspective
on this theorem and its application to quantum gauge theories through
the formalism of Hilbert subspaces of locally convex vector spaces
\ci{Sch1,Sch2,Tho}. Secondly, the `quantum reduction map' from the
unphysical state space to the physical \Hs\ turns out to map coherent
states for the gauge field $A$ into Hall's coherent states \ci{H1},
parametrized by the Wilson loop $\CW(A)\in K$.

Let us now turn to the details. In section 2 we explain the
application of \MW\ reduction to Yang--Mills theory. In section 3 we
take the first steps towards quantizing this procedure. Section 4
contains an intermezzo on Hilbert subspaces and measures on
infinite-dimensional spaces.  In section 5 we adapt and apply this
material to Yang--Mills theory.  In section 6 we review Hall's
coherent states, placing them in a general context.  Finally, section
7 outlines how the various threads come together.
\section{\hspace{-4mm}.\hspace{2mm}MARSDEN-WEINSTEIN REDUCTION IN GAUGE THEORY}
The starting point is a (strongly) symplectic manifold $S$,
mathematically representing the phase space of the system, with
associated Poisson bracket.  A Lie group $\CG$ is supposed to act on
$S$ in strongly Hamiltonian fashion. This means two things: firstly,
the action is canonical, in that the Poisson brackets are preserved,
and secondly, the action is infinitesimally generated in the following
sense.  For each $X$ in the Lie algebra $\g$ of $\CG$ there is a
function $J_X\in\cin(S)$ such that $\{J_X,f\}=\xi_X f$, where $\xi_X
f(\sg)=d f(\Exp(tX)\sg)/dt(t=0)$. The map $X\mapsto J_X$ is
automatically linear, defining a map $J:S\raw\g^*$ by $\la J(\sg),X\ra
=J_X(\sg)$. It is then required that this map be equivariant with
respect to the $\CG$-action on $S$ and the coadjoint action on $\g^*$.
Equivalently, $J$ is a Poisson map relative to the Lie--Poisson
structure on $\g^*$. One calls $J$ a momentum map for the given
$\CG$-action on $S$.  The \MW\ quotient \be S^0=J\inv(0)/\CG \ee is
then a symplectic manifold, provided that the $\CG$-action on $S$ is
proper and free. See \ci{AM,MT} for details.

This formalism has two applications to physics, which are identical in
the underlying mathematics, but have a different physical
interpretation.  In the oldest one, $\CG$ is the symmetry group of
some Hamiltonian $h$ on $S$.  Noether's theorem then states that each
$J_X$ is conserved under the flow generated by $h$, which quotients to
a well-defined function $h^0$ on the reduced phase space $S^0$. One
then attempts to solve the equations of motion on $S$ by finding the
flow of $h^0$ on $S^0$, and subsequently constructing a preimage of
this flow on $S$.  In a more modern application, $\CG$ is a gauge
group, and $S$ is the phase space of unconstrained degrees of freedom
of a gauge theory.  The reduced space $S^0$ is then the true phase
space of physical degrees of freedom of the system. This application
of symplectic reduction was first considered by J. Marsden and his
school; see \ci{Arms}. Typically, the constraint manifold $J\inv (0)$
consists of those fields that satisfy Gauss's law.

To apply this to \YM\ theory on the circle $\T$, we take the
configuration space of connections on the trivial principal $K$-bundle
$P=\T\x K$ to be the real \Hs\ $\CA=L^2(\T,\k)$, where $\k$ is the Lie
algebra of the compact structure group $K$. Identifying $\k$ with its
dual $\k^*$ through the choice of an invariant inner product, the
phase space $T^*\CA$ may be identified with the complex \Hs\ \be
S=L^2(\T,\k_{\C}) \ll{S} \ee of complexified $L^2$-connections. The
gauge group $\CG$ of this theory is the real Hilbert manifold
$\H_1(\T,K)$, consisting of all based loops $g\in C(\T,K)$ whose
(weak) derivative $\dot{g}=g\inv dg/dt$ lies in $L^2(\T,\k)$. We write
$Z=A+\half iE$.

The action \be g:Z\mapsto Z^g=\Ad(g)Z+ g dg\inv \ee of $\CG\ni g$ on
$S$, which is the pullback of the usual $\CG$-action on $\CA$, may be
shown to be smooth, proper, and free \ci{Raj2}, as well as strongly
Hamiltonian, with \mom\ $J$ \ci{LW,MT}.  Hence the \MW\ quotient $S^0$
is a symplectic manifold, which may be explicitly computed to be
canonically isomorphic to the cotangent bundle $T^*K$.  To understand
why, it is convenient to introduce the complexification $K_{\C}$ of
$K$. This is a Lie group whose Lie algebra is $\k_{\C}$ (the
complexification of $\k$), and which is diffeomorphic to $T^*K$
\ci{H3}.  The pertinent diffeomorphism equips $K_{\C}$ with a
symplectic structure, borrowed from the canonical one on $T^*K$. Now
recall the Wilson loop \be \CW(A)=P\,\Exp\left( -\oint_{\T}
A\right),\ll{wil} \ee seen as a map $\CW:\CA\raw K$.  Its
complexification $\CW_{\C}:S\raw K_{\C}$ is well-defined, and one may
prove \ci{MT}:
\begin{Theorem}
The map $\CW_{\C}$, restricted to $J\inv (0)$ is $\CG$-invariant, and
quotients to a symplectomorphism from $J\inv(0)/\CG$ to $K_{\C}$.
\end{Theorem}
The reduction to $T^*K$ was first found in \ci{Raj1}, and proved
rigorously in \ci{Raj2}.
\section{\hspace{-4mm}.\hspace{2mm}QUANTIZED MARSDEN--WEINSTEIN REDUCTION}
We now wish to quantize the classical setup. Traditionally, physicists
have used Dirac's method of constrained quantization, in which firstly
the unconstrained phase space $S$ is quantized into a \Hs\ $S$,
secondly the canonical $\CG$-action on $S$ is quantized into a unitary
\rep\ $U(\CG)$ on $\H$, and thirdly, the two-step classical reduction
procedure \be S\raw J\inv(0)\raw J\inv(0)/\CG \ee is replaced by the
single step of forming the physical subspace $\H_D$ of vectors in $\H$
that are invariant under all $U(g)$, $g\in\CG$. The physical inner
product on $\H_D$ is the one inherited from $\H$.  When $\CG$ is
connected, this is equivalent to saying that $\H_D$ is the subspace of
$\H$ that is annihilated by all quantized constraints $dU(X)$,
$X\in\g$. Hence Dirac quantized the first step of the classical
reduction procedure.

This method leads to severe mathematical difficulties, mostly because
$\H_D$ may well be empty, even when $S^0$ is not. While Dirac was
right in quantizing only half of the classical constraint algorithm,
in our opinion he misidentified which half. It turns out that a
mathematically correct theory is possible if one quantizes the second
step of quotienting by the $\CG$-action.  Namely, as explained in
detail in our talk at the 1995 Bia\l owie\.{z}a Workshop \ci{bial}
(also cf.\ \ci{MT}), the mathematical analogy between symplectic
reduction and induced \rep s of \ca s eventually leads to the
following algorithm of constrained quantization.

Firstly, the classical constraints define a positive definite
quadratic form $(\, ,\, )_0$ that is defined on a dense subspace
$\CD\subset\H$ (the passage from $\H$ to $\CD$ is a purely technical
functional-analytic matter, which should not be compared with the
passage from $S$ to the constraint manifold $J\inv(0)$ in the
classical theory). Secondly, writing $\CN$ for the null space of the
form $(\, ,\, )_0$, the physical \Hs\ $\H^0$ is the completion of
$\CD/\CN$ in the inner product inherited from $(\, ,\, )_0$ (and not
from the original inner product $(\, ,\, )$), which is positive
definite on $\CD/\CN$ precisely because its null vectors have been
thrown out.  A weak observable is an operator $A$ on $\H$ that leaves
$\CD$ stable and satisfies \be (\Ps,A\Ph)_0=(A^*\Ps,\Ph)_0 . \ll{obs}
\ee This property implies that $A\CN\subseteq\CN$, so that $A$ defines
an operator $A^0$ on the quotient $\CD/\CN$. This operator (which may
be extended to all of $\H^0$ under a suitable boundedness assumption)
is the physical observable defined by $A$.  More precisely, when
$V:\CD\raw \CD/\CN\subseteq\H^0$ is the canonical projection, one has
\be VA\Ps=A^0V\Ps \ll{qrm} \ee for all $\Ps\in\CD$. We call $V$ the
quantum reduction map. By definition of the inner product $(\, ,\,
)^0$ in $\H^0$, it satisfies \be (V\Ps,V\Ph)^0=(\Ps,\Ph)_0. \ll{Vp}
\ee It follows that one need not work with the abstract definition of
$\H^0$; in particular, it is not necessary to compute the null space
$\CN$. Given any \Hs\ $\til{\H}^0$ and a linear map $V:\CD\raw
\til{\H}^0$ which satisfies (\ref{Vp}) and has dense range, one may
define $A^0$ as an operator on $\til{\H}^0$ by means of (\ref{qrm}),
defining the physical theory on $\til{\H}^0$ rather than on
$\H^0$. The corresponding \rep\ $A\mapsto A^0$ of the algebra of all
weak observables is equivalent to the one originally defined on
$\H^0$.

In general, the quadratic form $(\, ,\, )_0$ is not even closable. In
the rare case that it is bounded, so that one may put $\CD=\H$, one
obtains $\CN$ as a closed subspace of $\H$, and $\H^0$ is isomorphic
to $\H_D=\CN^{\perp}$. This occurs when $\CG$ is compact, in which
case one finds \ci{bial,MT} \be (\Ps,\Ph)_0=\int_{\CG} dg\,
(\Ps,U(g)\Ph), \ll{fip} \ee where $dg$ is the normalized Haar measure
on $\CG$. One may bring the integration inside the inner product, and
using the expression $p_{\mbox{\tiny id}}=\int_{\CG} dg\, U(g)$ for
the projection on the trivial sub\rep\ of a compact group yields \be
(\Ps,\Ph)_0=(p_{\mbox{\tiny id}}\Ps,p_{\mbox{\tiny id}}\Ph).  \ee
Hence $\CN=(p_{\mbox{\tiny id}}\H)^{\perp}$, so that \be
\H^0=p_{\mbox{\tiny id}}\H=\H_D.  \ee This case is physically relevant
to gauge theories on a finite lattice, where the $\CG$-integration has
indeed been standard practice from the start.

However, when $\CG$ is noncompact but still locally compact, one
generically finds that the form (\ref{fip}) is only defined on a
proper dense subspace of $\H$, so that the general procedure just
described has to be followed. This is a typical situation where
Dirac's method breaks down but the improved method still works. Let us
note that for each $\ch\in \CG$ the operator $U(\ch)$ is a weak
observable, since (\ref{obs}) holds; in fact, assuming that the Haar
measure is right-invariant, both sides of (\ref{obs}) are equal to \be
(\Ps,U(\ch)\Ph)_0=(\Ps,\Ph)_0. \ll{gi} \ee This implies that the
physical observable defined by $U(\ch)$ is simply $U(\ch)^0=\I$ (the
unit operator on the physical state space $\H^0$). This property
guarantees the gauge invarince of the physical theory.

 A gauge group (in the continuum) is not even locally compact, so in
the absence of a Haar measure it is not clear what (\ref{fip}) should
mean. Before turning to this problem, let us write down the relevant
data for Yang--Mills theory on the circle \ci{LW,MT}. Since the
classical phase space (\ref{S}) is a \Hs, its quantization is the
bosonic Fock space $\H=\exp(S)$.  This space contains ``exponential
vectors'' $|Z\ra$ that are parametrized by $Z\in S$ and defined by
\ci{MT} \be |Z\ra =\sum_{l=0}^{\infty}\frac{\ot^l
Z}{\sqrt{l!}}=\Om+Z+\frac{Z\ot Z}{\sqrt{2!}}+\cdots. \ll{defExpv} \ee
The square-roots are explained by the property \be (\la
W|,|Z\ra)_{\exp(S)}=\exp(W,Z)_{S}. \ll{expip} \ee In the case at hand,
the general class of \rep s of gauge groups considered in \ci{GGV,AHK}
may be written as (henceforth omitting the suffix $S$) \be
U(g)|Z\ra=e^{-\half \n\dot{g}\n^2+(\dot{g},Z)}|Z^g\ra. \ll{GGVrep} \ee
In other words, the vector $|Z\ra$ transforms classically, up to a
prefactor that guarantees unitarity (see \ci{HD} for a non-unitary
approach to the reduction problem). Various arguments indicate that
$U(\CG)$ is indeed the correct quantization of the classical
$\CG$-action on $S$ \ci{Dim,LW,MT}.
\section{\hspace{-4mm}.\hspace{2mm} HILBERT SUBSPACES AND WIENER MEASURE}
We are now going to make sense of the expression (\ref{fip}) for our
gauge group $\CG$ of \YM\ theory on a circle (defined below
(\ref{S})). It turns out that, instead of using the non-existent
invariant Haar measure on $\CG$, we can define the form $(\, ,\, )_0$
in terms of a Gaussian measure $\mu$ on a certain completion $\CG^-$
of $\CG$. The special nature of the gauge group $\CG$ as a subgroup of
$\CG^-$ lies in the fact that $\mu$ is quasi-invariant under
translations by $\ch\in\CG^-$ iff $\ch$ lies in $\CG$. This property
will guarantee that (\ref{gi}) holds, implying gauge invariance of the
physical theory.

It is enlightening to present the general mathematical context of this
subtle phenomenon.  The following concept was introduced by Laurent
Schwartz \ci{Sch1,Sch2}. A Hilbert subspace of a topological vector
space $\CV$ is a \Hs\ $\H$ with continuous linear injection
$\H\hookrightarrow\CV$.  In other words, $\H$ is a continuously
embedded subspace of\/ $\CV$.  The Riesz--Fischer theorem then leads
to an antilinear map $\th\rat\til{\th}$ from $\CV^*$ to $\H$ (and
hence to $\CV$), defined by the property $\th(w)=(\til{\th},w)$ for
all $w\in\H$. One obtains a positive sesquilinear form $Q$ on $\CV^*$
by \be Q(\th,\et)=(\til{\et},\til{\th}). \ll{defQschwartz} \ee

For a simple example, consider the case that $\CV=\CV^*=\H\oplus\CK$
is itself a \Hs, with the obvious embedding of $\H$. Then
$\til{\th}=p_{\H}\th$ and $Q(\th,\et)=(\th,p_{\H}\et)$, where $p_{\H}$
is the orthogonal projection onto $\H$.

Now suppose that $\CV$ carries a Radon measure $\mu$ whose Fourier
transform is given by \be \int_{\CV}d\mu(v)\, e^{i\th(v)}=e^{-\half
Q(\th,\th)}, \ll{GCc1} \ee where $\th\in\CV^*$.  A measure with this
property is called Gaussian, with covariance $Q$, and is uniquely
determined by its Fourier transform (\ref{GCc1}).

The general Cameron--Martin theorem \ci{Tho} describes the behaviour
of $\mu$ under translation.  Recall that two measures are equivalent
if they have the same null sets, and disjoint if their supports are
disjoint.
\begin{Theorem}\ll{schwartz}
Let $\H$ be a real Hilbert subspace of a quasi-complete locally convex
Hausdorff vector space $\CV$.
\begin{enumerate}
\item
The map $\til{\th}\rat\hat{\th}$ from $\til{\CV}^*$ to $L^2(\CV,\mu)$,
defined by $\hat{\th}(v)=\th(v)$, is well-defined and isometric, so
that it extends to an isometry $w\rat\hat{w}$ from $\H$ to
$L^2(\CV,\mu)$.  For each $w\in\H$ this defines $\hat{w}$ as an
element of $L^2(\CV,\mu)${\rm ;} we write $(w,v)=\hat{w}(v)$, which
makes sense for almost all $v\in\CV$ with respect to $\mu$.
\item
The translate of $\mu$ by $w\in\CV$ is disjoint from $\mu$ when
$w\notin\H$, and equivalent to $\mu$ when $w\in\H$, with
Radon--Nikodym derivative \be d\mu(v+w)=e^{-\half
(w,w)-(w,v)}d\mu(v). \ll{CM} \ee
\end{enumerate}
\end{Theorem}

In the example $\CV=\H\oplus\CK$ considered above, assume that
$\H\simeq\R^n$ is finite-dimensional (for otherwise $\mu$ will not
exist). Then $\mu$ is simply the standard Gaussian measure on $\R^n$
supported on the hyperplane $\H$ in $\CV$, and the claims of the
theorem are obvious: a translation by $w\in\H$ doesn't change this
hyperplane, whereas translating by $w\notin\H$ moves it to a
hyperplane that is disjoint from $\H$.

Let us note that L.P. Gross's approach to measures on
infinite-dimensional vector spaces in terms of ``measurable norms''
(see \ci{Kuo}) may be seen in the above light, taking $\CV$ to be a
Banach space. The following example is a case in point (cf.\ \ci{Tho}).

Take \be \H=L^2([0,1],\R^n) \ee and \be \CV=C([0,1],\R^n)_0, \ee seen
as a Banach space in the supremum-norm; the suffix $\mbox{}_0$
indicates that elements of $\CV$ are (interpreted as) paths $A$ in
$\R^n$ that start at $0$ at $t=0$ (and continue until $t=1$). The
anti-derivative $\CP A(t)=\int_0^{t} ds\, A(s)$ embeds $\H$
continuously into $\CV$. The quadratic form $Q$ on the dual $\CV^*$,
consisting of the signed Radon measures on $[0,1]$ tensored with
$\R^n$, reads \be Q(\th,\et)=\sum_{i=1}^n\int_0^1\int_0^1 d\th_i(s)\,
d\et_i(t)\min(s,t).  \ee The Gaussian measure characterized by
(\ref{GCc1}) indeed exists, being nothing but the Wiener measure
$\mu_W$. Theorem \ref{schwartz} then reduces to the original
Cameron--Martin theorem \ci{CM}.  The Hilbert subspace $\CP
L^2([0,1],\R^n)$ of paths with finite energy is known as the
Cameron--Martin subspace of $C([0,1],\R^n)_0$, which plays a central
role in infinite-dimensional stochastic analysis \ci{Mal}.

Physicists sometimes write the Wiener measure as \be
d\mu_W[x(\cdot)]=N\left(\prod_{t=0}^1 dx(t)\right)\, e^{-\half
\int_0^1 \dot{x}^2}, \ll{wie} \ee where neither the infinite
normalization constant $N$ nor the infinite product makes mathematical
sense. Moreover, the $L^2$-norm of $\dot{x}$ is finite iff $x$ lies in
the Cameron--Martin subspace, which is unfortunately of
$\mu_W$-measure zero. Nonetheless, assuming that the ``Lebesgue
measure'' $dx=\prod_t dx(t)$ is translation-invariant, one may verify
(\ref{CM}) from (\ref{wie}). We will make heuristic use of (\ref{wie})
also in the next section.
\section{\hspace{-4mm}.\hspace{2mm} GAUGE INVARIANCE FROM THE 
CAMERON--MARTIN THEOREM}
In order to apply the results of the preceding section to \YM\ theory,
we should deal with the fact that our gauge group $\CG=\H_1(\T,K)$ is
not a linear space.  Nonetheless, it will play the role of the
Cameron--Martin subspace of $\CG^-=LK=C(\T,K)_e$, the group of all
based continuous loops in $K$, which is the analogue of $\CV$. This
time the inclusion $\CG\hookrightarrow LK$ continuously injects a
Hilbert manifold into a Banach manifold (the pertinent tangent spaces
are $T_e\CG=\g=\H_1(\T,\k)$, the Sobolev space of continuous loops in
$\k$ with derivative in $L^2$, and $T_e(LK)=C(\T,\k)_0$, the Banach
space of continuous loops in $\k$ with
supremum-norm, both classes of loops starting at $0$).
 
The passage from the space $C([0,1],\k)_0$ of the preceding section
(in which we put $\k\simeq\R^n$) to the based loop group $LK$ is
accomplished in two steps. Firstly, Ito's map
$\hat{\CI}:C([0,1],\k)_0\raw C([0,1],K)_e$ is defined by \ci{Fre,MM}
\be \hat{\CI}_X(t)=\lim_{N\raw\infty} \prod_{n=0}^{N-1}\Exp\left [
X\left((1-\frac{n+1}{N})t\right)-X\left((1-\frac{n}{N})t\right)\right].
\ll{Ito2} \ee It can be shown that the limit exists for almost every
$X$ with respect to the Wiener measure $\mu_W$.  The image of $\mu_W$
under Ito's map is the Wiener measure $\mu^{CK}_W$ on $C([0,1],K)_e$.
Ito's map is a bijection up to null sets of $\mu_W$ and
$\mu_W^{CK}$. Restricted to $\CP L^2([0,1],\k)\subset C([0,1],\k)_0$,
we may write Ito's map in terms of an incomplete Wilson loop as
$\hat{\CI}\circ \CP=\hat{\CW}$, where $\hat{\CW}:L^2([0,1],\k)\raw
C([0,1],K)_e$ is defined by \be \hat{\CW} (A):t\mapsto P\,\Exp\left (
-\int_0^t ds\, A(s)\right); \ll{wil2} \ee cf.\ (\ref{wil}). This may
seem pointless, because $\CP L^2([0,1],\k)$ has zero Wiener measure,
but the point is that, using the technique of stochastic differential
equations \ci{Fre,MM,HD}, eq.\ (\ref{wil2}) may be extended from the
domain $L^2([0,1],\k)$ to the domain of generalized derivatives of
functions in $C([0,1],\k)$.

Secondly, the measure $\mu^{CK}_W$ on $C([0,1],K)_e$ is conditioned so
as to become supported on $LK$, yielding the Wiener measure
$\mu^{LK}_W$ on $LK$. The analogue of Theorem \ref{schwartz}.2
\ci{Fre,MM,Sad} then reads
\begin{Theorem}\ll{FMM}
The translate of $\mu_W^{LK}$ by $\ch\in LK$ is disjoint from
$\mu_W^{LK}$ when $\ch\notin\CG$, and equivalent to $\mu$ when
$\ch\in\CG$, with Radon--Nikodym derivative \be d\mu_W^{LK}(g\ch)=
e^{-\half \n \dot{\ch}\n^2-(\dot{g},\Ad(\ch)\dot{\ch})}
d\mu_W^{LK}(g), \ll{CMLH} \ee where the second term in the exponential
is defined as in Theorem \ref{schwartz}.1.
\end{Theorem}

We now return to the problem of making sense of the expression
(\ref{fip}) for \YM\ theory, or, more precisely, of defining a
quadratic form $(\,,\,)_0$ on some dense domain $\CD\subset\exp(\CA)$
that satisfies (\ref{gi}). They key heuristic fact is that
(\ref{GGVrep}) leads to the expression \be (\la W|,U(g)|Z\ra
)=e^{-\half\n\dot{g}\n^2} e^{(W,Z^g) +(\dot{g},Z)}.  \ll{matrixYM} \ee
Combined with (\ref{wie}) and (\ref{fip}), this expression motivates
the definition of $(\,,\,)_0$ on the exponential vectors
(\ref{defExpv}) by \be (\la W|,|Z\ra )_0 = \int_{LK}d\mu_W^{LK}(g)\,
e^{(W,Z^g) +(\dot{g},Z)}.  \ll{ripYM} \ee Here the expressions of the
type $(\dot{g},Z)$, which in (\ref{matrixYM}) were well-defined for
$g\in \H_1(S^1,K)$ as inner products, make sense for general $g\in LK$
by Theorem \ref{schwartz}.1.  The domain $\CE$ of finite linear
combinations of the coherent states (\ref{defExpv}) is dense in
$\exp(\CA)$; it may be shown that (\ref{ripYM}) may be consistently
extended to $\CE$ (this is nontrivial, because the coherent states are
overcomplete). Thus we define $(\,,\,)_0$ on $\CD=\CE$ by sesquilinear
extension of (\ref{ripYM}).

The rigorous justification of (\ref{ripYM}), which in itself has been
obtained by a heuristic argument, is that (\ref{gi}) is satisfied for
all $\ch\in\CG$; this is an easy consequence of Theorem \ref{FMM}. As
explained in section 3, this implies that the physical theory is gauge
invariant under the gauge group $\CG$.  The gauge group of the theory
is $\CG$ rather than the auxiliary device $LK$: the \rep\ $U$ given in
(\ref{GGVrep}) cannot even be extended from $\CG$ to $LK$, and even if
it could, Theorem \ref{FMM} would make it clear that the fundamental
property (\ref{gi}) only holds for $\CG$ rather than for all of $LK$.
\section{\hspace{-4mm}.\hspace{2mm} HALL'S  COHERENT STATES}  Now that
we have found the quadratic form $(\,,\,)_0$ for \YM\ theory on a
circle, we may try to compute the physical \Hs\ $\H^0$ and the
associated \rep\ of the weak physical observables; see section 3. Here
Hall's coherent states turn out to play a crucial role. Let us first,
however, review the general notion of a coherent state \ci{KS,MT}.
\begin{Definition}\ll{defcs}
Given a manifold $S$, a subset $I\subset\R\backslash \{0\}$ having $0$
as an accumulation point, and a family $\{\Hh\}_{\hbar\in I}$ of \Hs
s, a system of coherent states is a collection
$\{\Ps_{\hbar}^{\sg}\in\Hh\}_{\hbar\in I}^{\sg\in S}$ of unit vectors,
along with a set $\{\mu_{\hbar}\}_{\hbar\in I}$ of Radon measures on
$S$, such that
\begin{enumerate}
\item
for each $\hbar\in I$ and all $\Ps\in\Hh$ one has the completeness
property \be \int_S d\mu_{\hbar}(\sg)\, |(\Ps_{\hbar}^{\sg},\Ps)|^2=1;
\ll{qhnorm} \ee
\item
for each $\hbar\in I$ the map from $S$ to the projective space $\PHh$
defined by projecting $\sg\mapsto \Ps_{\hbar}^{\sg}$ is continuous;
\item
for fixed $\rh$ and $\sg$ the function $\hbar\rat
|(\Ps_{\hbar}^{\sg},\Ps_{\hbar}^{\rh})|^2$ is continuous, with
classical limit \be \lim_{\hbar\raw 0}
|(\Ps_{\hbar}^{\sg},\Ps_{\hbar}^{\rh})|^2=\dl_{\rh\sg}. \ll{prsdlrs}
\ee
\end{enumerate}
\end{Definition}
 
For example, one may take $S$ to be a finite-dimensional \Hs\
$S\simeq\C^n$, $I=\R_{>0}$, $\Hh=\exp(S)$ (independent of $\hbar$),
$\mu_{\hbar}$ equal to $(2\pi\hbar)^{-n}$ times Lebesgue measure, and
\be \mbox{}_{\mbox{\tiny Fock}}\Ps_{\hbar}^{\sg}=e^{-\half
(\sg,\sg)/\hbar} |\sg/\sqrt{\hbar}\ra, \ll{nexp} \ee cf.\
(\ref{defExpv}). It is follows from (\ref{expip}) that \be
|(\mbox{}_{\mbox{\tiny Fock}}\Ps_{\hbar}^{\sg},\mbox{}_{\mbox{\tiny
Fock}}\Ps_{\hbar}^{\rh})|^2=e^{-|\rh-\sg|^2/\hbar}, \ee which implies
all properties in Definition \ref{defcs}. In case that $S$ is an
infinite-dimensional \Hs, this discussion is still valid if all
reference to the measures $\mu_{\hbar}$ (and therefore condition
(\ref{qhnorm})) is omitted. Hence the exponential vectors
(\ref{expip}) are essentially coherent states with $\hbar=1$, up to
normalization; the normalized coherent states are given by
(\ref{nexp}).

Hall's coherent states were introduced in \ci{H1}, but their
interpretation as coherent states satisfying Definition \ref{defcs}
only became clear from \ci{H3,H4}.  Their definition involves the
fundamental solution $\rh(x,t)$ of the heat equation
$df/dt-\half\Delta_K f=0$ on a compact connected Lie group $K$, as
well as the fundamental solution $\rh_{\C}(\sg,t)$ of the heat
equation $df/dt-\half\Delta_{K_{\C}} f=0$ on $K_{\C}$; here $\Dl_K$
and $\Delta_{K_{\C}}$ are the Laplacians on $K$ and on $K_{\C}$,
respectively; cf.\ \ci{H1}.  Hall proved in \ci{H1} that $\rh$ may be
analytically continued from $K$ to $K_{\C}$; we call this continuation
$\rh^{\C}$.
\begin{Definition}\ll{Hall}
Let $K$ be a compact connected Lie group. In the context of Definition
\ref{defcs} one takes $S=K_{\C}$ (cf.\ section 2), $I=\R_{>0}$,
$\Hh=L^2(K)$ (defined with respect to the normalized Haar measure, and
independent of $\hbar$), $\mu_{\hbar}$ as defined through its
Radon--Nikodym derivative with respect to the Haar measure $d\sg$ on
$K_{\C}$ by \be d\mu_{\hbar}(\sg)=d\sg\int_K dk\,
\rh_{\C}(k\inv\sg,\hbar), \ee and finally \be \mbox{}_{\mbox{\tiny
Hall}}\Ps_{\hbar}^{\sg}:k\mapsto N_{\hbar} \rh^{\C}(k\inv \sg,\hbar),
\ee where $N_{\hbar}$ is a certain ($\sg$-independent) normalization
constant guaranteeing that $\mbox{}_{\mbox{\tiny
Hall}}\Ps_{\hbar}^{\sg}$ is a unit vector in $L^2(K)$.
\end{Definition}

It is possible to use this definition for $K=\R^n$, so that
$K_{\C}=\C^n$; using the explicit expressions for the heat kernels on
$\R^n$ and $\C^n$, one recovers the coherent states (\ref{nexp}) in
the Bargmann--Fock \rep. See \ci{H3,H4}.
\section{\hspace{-4mm}.\hspace{2mm}PUNCH LINE}
All threads now come together.
\begin{Theorem}\ll{pl}
In the discussion following (\ref{Vp}) we take $(\, ,\,)_0$ and $\CD$
as defined in (\ref{ripYM}) and subsequent text, and
$\til{\H}^0=L^2(K)$.
\begin{itemize}
\item
The quantum reduction map $V:\CD\raw L^2(K)$, given by linear
extension of \be V\mbox{}_{\mbox{\tiny
Fock}}\Ps_{1}^{Z}=\mbox{}_{\mbox{\tiny
Hall}}\Ps_{\half}^{\CW_{\C}(Z)}, \ll{int1} \ee satisfies
(\ref{Vp}). Here $Z\in S=L^2(\T,\k_{\C})$ (cf.\ (\ref{S})), and the
complexified Wilson loop $\CW_{\C}(Z)\in K_{\C}$ is defined below
(\ref{wil}).
\item
For each $\ch\in\CG$ one has \be VU(\ch)\, \mbox{}_{\mbox{\tiny
Fock}}\Ps_{1}^{Z}=\mbox{}_{\mbox{\tiny
Hall}}\Ps_{\half}^{\CW_{\C}(Z)}. \ll{gial} \ee
\item
Let $f\in\cin(K)$, defining the function $\CW_f: A\mapsto f(\CW(A))$
on $S$. These functions can be quantized by certain operators
$\CQ(\CW_f)$ on $\exp(S)$; see \ci{MT}. One then has \be V\CQ(\CW_f)\,
\mbox{}_{\mbox{\tiny Fock}}\Ps_{1}^{Z}=f\mbox{}_{\mbox{\tiny
Hall}}\Ps_{\half}^{\CW_{\C}(Z)}. \ll{int2} \ee
\end{itemize}
Hence the physical \Hs\ of \YM\ theory on the circle may be identified
with $L^2(K)$, on which the gauge group acts trivially, and functions
of the Wilson loop act as multiplication operators.
\end{Theorem}

The calculations leading to (\ref{int1}) and (\ref{int1}) are
presented in \ci{Wren1,Wren2,MT}.  The proof that Hall's coherent
states are total in $L^2(K)$ is in \ci{H1}; this is necessary in view
of the discussion after (\ref{Vp}). Finally, (\ref{gial}) is a direct
consequence of (\ref{gi}); see the end of the previous section.

There is a statement similar to (\ref{int2}) in which the physical
Hamiltonian $-\half\Delta_K$, is derived from the Hamiltonian of the
unconstrained theory; see \ci{Wren2,Wren3}. However, the latter is an
extremely involved operator, and it seems that the techniques
developed in \ci{HD} are more suitable to deal with it than ours.

It may be argued that we have quantized a rather simple model in an
incredibly complicated way.  However, in higher-dimensional gauge
theories the reduced phase space is not known explicitly, and one must
deal with the constraints in some way. We hope that both our general
method of constrained quantization and the special techniques used to
apply this method to \YM\ theory on a circle can be generalized to
higher dimensions. The essential task will be to realize the gauge
group as a generalized Cameron--Martin subspace of some enlargement of
it, along with a suitable generalization of Theorem \ref{FMM} that
eventually guarantees the gauge invariance of the physical quantum
theory. It seems to us that the probabilistic literature offers some
hope for this to be possible.
\section*{\hspace{-2mm}ACKNOWLEDGEMENTS}
 N.P. Landsman is supported by a fellowship from the Royal Netherlands
Academy of Arts and Sciences (KNAW). He is grateful to Erik Thomas for
explaining the Hilbert subspace formalism to him. Both authors thank
Brian Hall for educating them on the Cameron--Martin subspace and
related matters.

\end{document}